\documentclass[12pt,twoside]{article}
\usepackage{epsf,epsfig,rotating}
\usepackage{graphicx}
\def\21{$SU(2) \otimes U(1) $}
\def\ne{\hbox{$\nu_e$ }}
\def\nm{\hbox{$\nu_\mu$ }}

\def\lsim{\raise0.3ex\hbox{$\;<$\kern-0.75em\raise-1.1ex\hbox{$\sim\;$}}}
\def\gsim{\raise0.3ex\hbox{$\;>$\kern-0.75em\raise-1.1ex\hbox{$\sim\;$}}} 

\newcommand{\mx}{\left[\begin{array}}
\newcommand{\finmx}{\end{array}\right]} 
\newcommand{\mxp}{\left(\begin{array}} 
\newcommand{\finmxp}{\end{array}\right)} 
\def\beq{\begin{equation}}
\def\eeq{\end{equation}}
\def\bea{\begin{eqnarray}}
\def\eea{\end{eqnarray}}

\def\mathbf#1{\hbox{\bf #1}}
\def\textrm#1{\hbox{#1}}
\def\lsim{\raise0.3ex\hbox{$\;<$\kern-0.75em\raise-1.1ex\hbox{$\sim\;$}}}
\def\gsim{\raise0.3ex\hbox{$\;>$\kern-0.75em\raise-1.1ex\hbox{$\sim\;$}}}
\newcommand {\ignore}[1]{}
\def\ap{\approx}

\begin{document}
\begin{titlepage}
\bibliographystyle{unsrt}
\hfill hep-ph/0001039\\
\hspace*{1cm}\hfill CERN-TH/2000-002\\
\hspace*{1cm}\hfill FTUV/99-71\\
\hspace*{1cm}\hfill IFIC/99-74\\
\vskip 0.3cm    
\begin{center}
{\Large Supernova Bounds on Majoron-emitting decays of light neutrinos}
\vskip 0.5cm
{\large M. Kachelrie{\ss}$^{1,2}$, R. Tom\`as$^2$ and J.~W.F. Valle$^2$}\\
\vskip0.3cm
$^1$ TH Division, CERN, CH-1211 Geneva 23 \\ \vskip0.2cm
$^2$ Institut de F\'{\i}sica Corpuscular -- C.S.I.C. \\
Departament de F\'{\i}sica Te\`orica -- Univ. de  Val\`encia \\
Edifici Instituts de Paterna -- Apartat de Correus 2085 --  
46071, Val\`encia \\
http://neutrinos.uv.es 
\end{center}
\date{\today}  
\vskip 1cm

\abstract{Neutrino masses arising from the spontaneous violation of
  ungauged lepton--number are accompanied by a physical Goldstone
  boson, generically called Majoron.  In the high--density supernova
  medium the effects of Majoron--emitting neutrino decays are
  important even if they are suppressed \emph{in vacuo} by small
  neutrino masses and/or small off-diagonal couplings.  We reconsider the
  influence of these decays on the neutrino signal of supernovae in
  the light of recent Super--Kamiokande data on solar and atmospheric
  neutrinos. We find that majoron--neutrino coupling constants in the
  range $3 \times 10^{-7} \lsim g \lsim 2  \times 10^{-5}$ or $g \gsim 3
   \times 10^{-4}$ are excluded by the observation of SN
  1987A.  Then we discuss the potential of Superkamiokande and
  the Sudbury Neutrino Observatory to detect majoron neutrino
  interactions in the case of a future galactic supernova.  We find
  that these experiments could probe majoron neutrino interactions
  with improved sensitivity.}

\end{titlepage}

\section{Introduction}

The solar~\cite{solarexp99} and atmospheric~\cite{atmexp99} neutrino
problems have provided two milestones in the search for physics beyond
the Standard Model, giving strong
evidence for \ne and \nm conversions, respectively. Although 
flavour changing neutral current interactions
(FCNC)~\cite{Schechter:1980gr,Hall:1986dx,Bergmann:1998rg} and/or neutrino
decays~\cite{Valle:1983ua} may play an important r\^ole in the
interpretation of the
data~\cite{Barger:1999bg,Krastev:1997cp,Gonzalez-Garcia:1998hj}, 
we concentrate here on oscillations involving very small neutrino mass
splittings, because they provide the simplest
picture~\cite{valencia:1999,Barger:1999,atm98}. Moreover, reconciling
the LSND data \cite{Louis:1998qf} and/or hot dark matter suggests the
existence of a light sterile neutrino \cite{Peltoniemi:1993ss}.
Pending the confirmation of LSND results by future experiments, we
choose however to use in this work just the solar and atmospheric data
assuming the absence of sterile neutrinos, FCNC and/or of heavier neutrinos.

Within the simplest extensions of the Standard Model in which neutrino
masses are introduced by an explicit breaking of lepton--number via
the seesaw mechanism there are neutral-current-mediated neutrino decays
$\nu' \to 3 \nu$ \cite{Schechter:1980gr}, but these decays are extremely slow
for the neutrino masses of interest for us here.
If neutrino masses arise from the spontaneous violation of ungauged
lepton number, the corresponding Goldstone boson
\cite{Chikashige:1981ui} brings in the possibility of new, potentially
faster, 2-body neutrino decays~\cite{Valle:1983ua},
\begin{equation}
\label{eq:J}
\nu'  \to  \nu + J  \:.
\end{equation}
However it is well-known that in the simplest well--motivated majoron
models the neutrino mass and coupling matrices are proportional, so
that decays are highly suppressed \emph{in vacuo}%
\footnote{This suppression can be avoided in some models due to a
  judicious choice of the quantum numbers
  \protect~\cite{Valle:1983ua,Valle:1991pk}.
  In such case the effects discussed here, although important, are not
  essential.} \cite{Schechter:1982cv}. 
As a result, neutrino decays become cosmologically and astrophysically
irrelevant in these models.  Moreover, for the small masses indicated
by solar and atmospheric neutrino data, the annihilation channels $\nu'
+\nu' \to J + J$ are also negligible.

The purpose of the present work is to discuss the possible impact of
neutrino-majoron interactions on the supernova (SN) neutrino signal.
Together with the early Universe SN are the only site where
neutrinos are in thermal equilibrium and so abundant that
neutrino-neutrino interactions become important. Therefore, SN physics
has been one of the main tools to derive limits on neutrino parameters
in majoron models \cite{ko82,be89,ch88,ch90}.

The arguments used in the literature can be divided mainly into two
classes: The first uses the fact that majoron interactions violate
lepton number (like in $\nu \to \bar\nu+J$ or $\nu+\nu \to \bar\nu+
\bar\nu$) and, thereby, can reduce the trapped electron lepton number
fraction $Y_{L_e}$ in the SN core \cite{ko82,be89}.  Since in several
SN simulations it was found that a minimum value of $Y_{L_e}\gsim
0.38$ is needed in order to have a successful bounce shock, the
strength of the lepton number violating interactions can be
correspondingly restricted.  Unfortunately, the older works
\cite{ko82} using this argument did not take into account the
derivative nature of the coupling of a Goldstone boson calculating the
scattering cross sections. Moreover, an approximation valid only in
vacuum was used for the neutrino decay rates. In ref.~\cite{be89} the
correct medium decay rates were used for the first time, giving the
stringent limit $g_{ee} \lsim 5 \times 10^{-6}$.

The second argument used is a ``classical'' energy loss argument: the
observed neutrino signal of SN 1987A should not be shortened too much
by additional majoron emission \cite{be89,ch88,ch90}.  The most
comprehensive work along this line is Ref.~\cite{ch90} which also uses
the correct derivative coupling of the majoron.  The limits found
there depend on the value of the $(B-L)$ breaking scale. For example
the range $5 \times 10^{-7} \lsim g \lsim 6 \times 10^{-5}$ is excluded
for $v=20$~GeV while $1 \times 10^{-5} \lsim g \lsim 7 \times 10^{-5}$ is
excluded for $v=500$~GeV.  The tau-neutrino masses considered in
Ref.~\cite{ch90} are, compared to the values currently discussed,
rather large, 100~eV $\lsim m_{\nu_\tau} \lsim 30$~MeV. It is
therefore of certain interest to extend their discussion to lower
neutrino masses presently indicated by the solutions to the solar and
atmospheric neutrino anomalies.

These astrophysical limits should be confronted with the available
laboratory constraints. While there is a stringent limit from
$\beta\beta$ experiments \cite{exp1}, 
\beq
 \sum_{i,j} g_{ij} U_{ei}U_{ej} \lsim 3  \times 10^{-5} \,, 
\eeq
the limits from pion \cite{exp2} and kaon decays \cite{exp3} are
rather weak,
\beq   \label{pi}
\sum_{l=e,\mu,\tau} g_{el}^2 \lsim  3  \times 10^{-5} \quad{\rm and}\quad
\sum_{l=e,\mu,\tau} g_{\mu l}^2 \lsim  2.4  \times 10^{-4} \,.
\eeq
Note also that individual
couplings $g_{ij}$ could be larger than the limit from $\beta\beta$
experiments due to possible cancellations~\cite{Wolfenstein:1981kw}.

The main purpose of the present work is the study of the impact of
neutrino-majoron interactions on the {\em observable\/} neutrino
signal of a supernova. Since decays of the type $ \bar\nu_e
\to\nu_l+J$ reduce the $ \bar\nu_e$ flux, a limit on the neutrino
majoron coupling constants can be derived from the observed signal of
SN 1987A. In so--doing one must include in the analysis the fact that
massive neutrinos may oscillate on their way from the SN envelope to
the detector. We do that and present the excluded regions for the
three currently discussed solutions of the solar neutrino problem
~\cite{valencia:1999,Barger:1999}.  Then we estimate the discovery
potential of new experiments like Superkamiokande (SK) and the Sudbury
Neutrino Observatory (SNO) in the case of a future galactic supernova.  We
find that these experiments could probe majoron neutrino coupling
constants $g$ down to $g \gsim \mathrm{few} \times 10^{-5}$.

\section{Matter effect on neutrino-majoron interactions}

We consider the simplest class of models in which neutrinos acquire
mass from the spontaneous violation of ungauged lepton number
\cite{Chikashige:1981ui}. In this case it is well-known that the
massless Goldstone boson $J$ -- the majoron -- couples diagonally to
the mass-eigenstate neutrinos $\nu_i$ to a very good
approximation~\cite{Schechter:1982cv}. In other words, after rotation
from the weak basis $\nu_\alpha$ through $\theta_0$ angle(s) the
original coupling matrix $g_{\alpha\beta}$ transforms into
\beq
g_{ij} \approx \delta_{ij} g_i \,.
\eeq
We denote by $\nu_i^{(h_i)}$ the 4-spinor describing the majorana
neutrino with mass $m_i$ and helicity $h=\pm 1$.

In this section, we briefly review the effect of a thermal background
on the neutrino majoron interactions. It was first demonstrated by
Berezhiani and Vysotsky in Ref.~\cite{be87} that the effective mass
induced by the interactions of neutrinos with background matter can
break the proportionality between the mass matrix $m_{ij}$ and the
coupling matrix $g_{ij}$ characteristic of the simplest majoron models
\emph{in vacuo} \cite{Schechter:1982cv}. A thermal background consists,
except in the early Universe, only of particles of the first
generation and, therefore distinguishes the electron flavour from the
other flavours.

In Ref.~\cite{be87}, the Lagrangian describing neutrino majoron
interaction in a thermal background was obtained in a relativistic
approximation similar to the usual treatment of neutrino oscillations
in matter. Later, the authors of Ref.~\cite{gi92} solved this problem
without using this approximation and confirmed the results of
Ref.~\cite{be87} in the appropriate limiting cases.  Here we are
concerned with SN neutrinos, which have typical energies around
$10-25$~MeV. Therefore, we will take advantage of the simpler
relativistic approximation and we will follow closely
Ref.~\cite{be87}.

The Hamiltonian $H_{\rm tot} \equiv H (x)$ describing the evolution of
neutrinos may be given as
\bea
 H_{\rm tot} &=& H_0 + H_{\rm med} + H_{\rm int} 
\\ &=&
   H_0 
   + \sum_{i,j}\sum_{h_i,h_j}   \bar\nu_i^{(h_i)}  V_{ij}^{h_i,h_j}  
                                           \nu_j^{(h_j)}  
   + g_{ij} \bar\nu_i^{(h_i)}  \gamma_5 \nu_j^{(h_j)}  J  \,,
\eea             
where the free Hamiltonian $H_0$ describes the propagation in vacuo,
$H_{\rm med}$ describes the effects of matter and $H_{\rm int}$ takes into
account the presence of neutrino-majoron interactions which may lead
to decays. 
Instead of using the eigenstates of the free Hamiltonian $H_0$ as
basis for perturbation theory, we will use the eigenstates of
$H_0 + H_{\rm med}$. Therefore, as a first step the Dirac equation, 
$(H_0+V)\nu_j^{(\pm)}=  i \partial_t\nu_j^{(\pm)}$ 
has to be solved\footnote{The majoron remains massless because the
  forward scattering amplitude of Goldstone bosons on matter
  vanishes.}.  
In the case of ultra-relativistic neutrinos, it is well-known that a
vector-like potential does not change the helicity of the neutrinos
--- its only effect is a rotation of the eigenstates of $H_0 + V$ 
with respect to the mass basis \cite{ma88}.
Therefore, the Dirac equation simplifies and reduces to the 
standard form known from neutrino oscillations,
\beq
 i\partial_t \nu_i^{(h)} = (H^{\rm rel}_{ij}+U_{i\alpha} V_{\alpha \beta} U^\dagger
_{\beta j})
\nu_j^{(h)}
\eeq
where $ H^{\rm rel}_{ij} \approx (p+m_i^2/(2p))\delta_{ij} $ and
$V_{\alpha\beta}$ is the potential matrix in the weak basis
\beq
V_{\alpha\beta} = \left(\begin{array}{ccc} V_C+V_N & 0 & 0\\ 0 & V_N &
  0 \\ 0 & 0 & V_N\end{array}
\right) \,.
\eeq
The potentials induced by the charged and neutral currents are $V_C =
\sqrt{2} h G_F n_B ( Y_e + Y_{\nu_e})$ and $V_N = \sqrt{2} h G_F n_B
\left( -\frac{1}{2}Y_N + Y_{\nu_e} \right)$, where $Y_i=(n_i-n_{\bar
  i})/n_B$ and $n_B$ is the baryon density. Finally,
$U$ is the mixing matrix relating mass and weak basis and
defined through $\nu_i = U_{i\alpha}\nu_ \alpha$. Diagonalizing
$H^{\rm rel}+U V U^\dagger $ gives the medium states
$\tilde{\nu}_i^{(h)}=\tilde{U}_{ij}^{(h)} \nu_{j}^{(h)}$.

In the case of a two-flavor neutrino system the mixing matrices, $U$ and
$\tilde{U}$, can be parametrized by $\theta_0$ and $\theta^{(h)}$
respectively. The diagonalization of
the Hamiltonian leads us to the following expression for the effective
mixing angle,
\beq
 \sin^2 2\theta^{(\pm)}(p) = 
 \frac{\sin^2 2\theta_0}{\sin^2 2\theta_0 + (\cos 2\theta_0-\xi^{(\pm)})^2 } 
 \:,
\eeq
where
\beq
 \xi^{(h)}= h \: \frac{\Delta_0}{2p\; G_F (Y_e+Y_{\nu_e})n_B} \:,
\eeq
and $\Delta_0=m_2^2-m_1^2$.
The effective masses in the medium are 
\beq
 m_{1,2}^{(h)\; 2} = -hp(2V_N+V_C)+\frac{1}{2}(m_1^2+m_2^2) 
                   \mp \frac{1}{2}\Delta^{(h)}  \:,
\eeq
where the upper sign is for $m_1$, the lower for $m_2$ and 
\beq
 \Delta^{(h)} = 
 \sqrt{(\Delta_0\cos 2\theta_0 + 2hpV_C)^2+(\Delta_0\sin 2\theta_0 )^2} \:.
\eeq

For typical
neutrino energies and densities near the neutrino--spheres, the parameter

\beq
 \xi =   6.53 \times 10^{-8}
         \frac{\Delta_0}{10^{-3}{\rm eV}^2} \; 
         \frac{10 {\rm MeV}}{p} \;
         \frac{10^{10}{\rm g/cm}^{-3}}{(Y_e+Y_{\nu_e})\rho} 
\eeq
is much smaller than one. This fact will give rise to the following
simplification, $\theta^{(+)}(p) \ap -\theta_0$ and
$\theta^{(-)}(p)\ap \pi/2 -\theta_0$, which, as it will be shown
later, will allow us to identify medium and weak interaction states.

In the three-flavor neutrino case the mixing matrix $U$ can be
parametrized as $U=U_{12}U_{13}U_{23} U_0$, where the matrices
$U_{ij}= U_{ij}(\theta_{ij})$ perform the rotation in the $ij$-plane
by the angle $\theta_{ij}$ and $U_0$ includes possible CP-violation
effects~\cite{Schechter:1980gr}. In the following we will assume for
simplicity CP conservation and $\theta_{13}=0$, the latter motivated
both by detailed fits of the atmospheric neutrino anomaly, but also by
the results of the Chooz experiment~\cite{Bemporad:1999}.  This
simplifies the mixing matrix to $\nu_i=U_{i\alpha}\nu_\alpha=
U_{12}U_{23}\nu_\alpha$ \cite{Schechter:1980} and we can make the
assignment, $\theta_{12}=\theta_\odot$ and $\theta_{23}=\theta_{\rm
  atm}$.  Notice that for light neutrinos near the neutrino spheres
the condition $|V_{\alpha\alpha}|\gg m_i^2/(2p)$ ($\xi\ll 1$) holds
and, since in the weak basis the potential is diagonal, the medium
states can be identified with the weak ones up to a rotation in the
$\nu_\mu-\nu_\tau$ subspace.  The expressions can be simplified by
choosing this arbitrary rotation angle to coincide with
$\theta_{23}$.  Then, only one angle, $\theta^{(h)}$, will be
required to connect medium and mass eigenstates, and one will be able
to \emph{recover} the two-flavor neutrino case,
\begin{equation}
\tilde{\nu}_i^{(h)}=\tilde{U}_{ij}^{(h)}\nu_j^{(h)} = \tilde{U}_{ij}
(\theta^{(h)})\nu_j^{(h)}=
\left\{ \begin{array}{lcl} 
 \tilde{\nu}_i^{(+)}&=&\tilde{U}_{ij}(-\theta_{12})~~\nu_j^{(+)}\\
 \tilde{\nu}_i^{(-)}&=&\tilde{U}_{ij}(\pi/2-\theta_{12})~\nu_j^{(-)}\\
\end{array} \right . \,.
\end{equation}
Thus the medium states can be identified with weak interaction states
according to Table \ref{states} and the coupling matrix
$\tilde{g}_{ij}$ in the medium basis can be approximated by the one in
the weak basis, $g_{\alpha \beta}$.

\begin{table}[!h]
\begin{center}

 \begin{tabular}{c|c|c}
 medium state & weak state & potential \\ \hline
$\tilde{\nu}_1^{(+)}$ & $\bar{\nu}_e$ &$ -(V_C+V_N)$\\
$\tilde{\nu}_2^{(+)}$ & $c_{23}\bar{\nu}_\mu+s_{23}\bar{\nu}_\tau$ &$ -V_N$\\
$\tilde{\nu}_3^{(+)}$ & $-s_{23}\bar{\nu}_\mu+c_{23}\bar{\nu}_\tau$ &$ -V_N$\\
$\tilde{\nu}_1^{(-)}$ & $c_{23}\nu_\mu+s_{23}\nu_\tau$ &$ V_N$\\
$\tilde{\nu}_2^{(-)}$ & $\nu_e$ &$ V_C+V_N$\\
$\tilde{\nu}_3^{(-)}$ & $-s_{23}\nu_\mu+c_{23}\nu_\tau$ &$ V_N$\\
 \end{tabular}
\end{center}
\caption{Medium eigenstates $\tilde{\nu}_i^{\pm}$, equivalent weak
eigenstates in the limit $\xi\ll 1$ and their potential energy 
$V=m^{(\pm)2}/(2p)$.}
\label{states}
\end{table}

Above, we have derived the transformation matrix
$U_{ij}(\theta^{(\pm)})$ between mass and medium eigenstates. The
relation between the corresponding coupling matrices \emph{in vacuo} $
g_{if}$ and \emph{in medium} $\tilde g_{if}$ is given by
\beq
 \tilde g_{if}^{h_i\to h_f} = \tilde{U}_{in}(\theta^{(h_i)})g_{nm}
\tilde{U}_{mf}^T(\theta^{(h_f)})
 =
 \left( \begin{array}{ccc}
         c_f c_i \,g_{11} + s_f s_i \,g_{22} & 
         s_i c_f \,g_{22} - c_i s_f \,g_{11} & 0\\ 
         c_i s_f \,g_{22} - s_i c_f \,g_{11} &
         s_f s_i \,g_{11} + c_f c_i \,g_{22} & 0\\
        0 & 0& g_{33}
       \end{array}\right) 
\eeq
with $c_{i,f}=\cos[\theta^{(\pm)}(p_{i,f})]$ and
$s_{i,f}=\sin[\theta^{(\pm)}(p_{i,f})]$.  Notice that the coupling
constant $g_{33}$ will only appear in decays not involving electron
neutrinos or anti--neutrinos, so that it will not be important for us.

Within our relativistic approximation, only helicity--flipping
neutrino decays $\nu_i^{\pm}\to\nu_j^{\mp}+J$ occur. For these decays,
the differential decay rate is
\beq  \label{dGamma}
 \frac{d\Gamma}{dp_f} = \frac{\tilde g_{if}^2}{8\pi}
  \left( \frac{p_i-p_f}{p_i^2} \right) 
  \left( \frac{m_i^{(\pm)2}}{2p_i} - \frac{m_f^{(\pm)2}}{2p_f} \right) \:.
\eeq
In the next section, we will apply this formula to supernova.  Additional
simplifications of the limit $\xi \ll 1$ are in this
case $m_{i,f}^{(\pm)2}/(2p_{i,f}) = V_{i,f}$ and that the coupling
constants do not depend on $p_{i,f}$. Hence, we can integrate
Eq.~(\ref{dGamma}) and obtain as total decay rate
\beq  \label{Gamma}
 \Gamma = \frac{\tilde g_{if}^2}{16\pi} (V_i-V_f) 
\eeq
and as average energy of the final neutrino $\langle E_f
\rangle=E_i/3$.  Obviously, only those decays are possible for which
$V_i-V_f>0$.

Besides enhancing the rates for neutrino decay, the dispersive effects
of the medium open also completely new decay channels of the majoron
into neutrinos \cite{be94}.  The majoron decays $J
\to\nu_i^{\pm}+\nu_j^{\pm}$ have the same matrix elements like the
neutrino decays $\nu_i^{\mp} \to\nu_j^{\pm}+J$ considered above. Their
total decay rate is \cite{be94}
\beq  \label{GammaJ}
 \Gamma = \frac{\tilde g_{ij}^2}{8\pi} (-V_i-V_j) \: S 
\eeq
and now those decays are possible for which $V_i+V_j<0$. 
The symmetry factor $S$ is $S=1/2$ if the two neutrinos 
are identical, and $S=1$ otherwise. 
Finally, we note that the neutrinos are emitted isotropically.

\section{Supernova neutrinos and neutrino majoron decays}

Here we first collect the available limits on majoron neutrino
coupling constants from the observation of the neutrino signal of SN
1987A and comment on their validity. Then, we discuss the sensitivity
of new experiments like SK or SNO to probe neutrino-majoron
interactions in the case of a future galactic supernova.

\subsection{Constraints from collapsing phase}

Massive stars become inevitably unstable at the end of their life,
when their iron core reaches the Chandrasekhar limit. The collapse of
the iron core is only intercepted when nuclear density, $\rho_0 \ap
3 \times 10^{14}$~g/cm$^3$, is reached. At this point, the implosion is
turned into an explosion: a shock wave forms at the edge of the core
and moves outward.  The strength of this bounce shock and its
successful propagation is extremely sensitive to the trapped electron
lepton fraction $Y_{L_e}=Y_e+Y_{\nu_e}$, attained by the core during
its infall. A successful SN explosion occurs only if at least 90\% of
the initial $Y_{L_e}$ is still present \cite{br89}, which translates
to $Y_L(t_{\rm bounce})\gsim 0.375$ \cite{ba89}.

We now use this requirement in order to derive a limit on majoron
decays. Such decays clearly change $Y_{L_e}$ either by two units
($\nu_e \to \bar\nu_{e} +J$) or by one ($\nu_e \to \bar\nu_{\mu} +J$).
Since the allowed change in $Y_L$ is small, we can still use the
profiles $Y_e(t)$ and $Y_{\nu_e}(t)$ from a ``standard'' SN simulation
\cite{br85}.  One can easily check from Table~1 that only the first of
the above decays takes place, since $Y_e+3/2Y_{\nu_e}< 1/2$.  Hence
the deleptonization rate is governed by
\beq \label{dY}
 \frac{dY_{L,{\rm decays}}}{dt}
                 = -2 \Gamma(\nu_e \to \bar\nu_{e} +J)  Y_{\nu_e} \,.
\eeq
We integrate Eq.~(\ref{dY}) numerically from $t_0$, the time when
neutrinos start to become trapped (at $\rho(t_0)\ap 5 \times
10^{11}$~g/cm$^3$), to the time of the bounce $t_{\rm bounce}$. Note
that in the numerical simulation \cite{br85} whose results we are
using $Y_L$ increases from $Y_L(t_0)\ap 0.37$ up to $Y_L(t_{\rm
  bounce})\ap 0.39$.  Requiring that $Y_L(t_{\rm bounce})> 0.375$
\cite{br89,ba89}, i.e.  $|\Delta Y_{L,{\rm decays}}|<0.015$, we obtain
\beq \label{l1}
 g_{ee} = g_{11}\cos^2\theta_0 + g_{22}\sin^2\theta_0 \lsim 2 \times 10^{-6}\,.
\eeq
Using the same argument, the limit $g_{ee}\lsim 5 \times 10^{-6}$ was
derived in Ref.~\cite{be89}.

This limit relies on numerical modeling of SN explosions, in
particular on the success of the explosion in specific models.
However, it is probably fair to say that current supernova models have
generally problems to produce successful explosions. Therefore,
Eq.~(\ref{l1}) cannot be viewed as a trustworthy limit as long we do
not have a better understanding of SN dynamics.

\subsection{Constraints from majoron luminosity}

Numerical computations \cite{Eb} of the total amount of binding energy
$E_b$ released in a supernova explosion yield, within a plausible
range of progenitor star masses and somewhat depending on the equation
of state used,
\beq
 E_b = (1.5-4.5) \times 10^{53}\: {\rm erg} \,.
\eeq
This range of values is confirmed by likelihood analysis of the
observed $ \bar\nu_e$ spectrum of SN 1987A under the hypothesis of
small mixing of $ \bar\nu_e$ with other neutrino flavours \cite{je96}.
Therefore, the parameter space of models which give rise to majoron
luminosity large enough that the observed $\bar\nu_e$ signal is
significantly shortened can be restricted. The most comprehensive
analysis of SN cooling and majoron emission using this argument was
given in Ref.~\cite{ch90}.

Our following analysis has three main differences compared to
Ref.~\cite{ch90}: First, we are interested in relatively light
neutrinos ($m^2\lsim {\rm eV}^2 \ll 2pV$) as suggested by the simplest
interpretation of data from solar and atmospheric neutrino
experiments.  Therefore, medium effects become important not only for
decays like $\nu \to\nu'+J$ but also for scattering processes like
$J+\nu \to J+\nu$. Second, we do not include the process $J+J \to J+J$
in the calculation of the majoron opacity. A discussion of why
self-scattering processes do not contribute to the opacity was given
in Ref.~\cite{di89} for the case of $\nu+\nu \to\nu+\nu$ scattering.
Note that trapping due to $J+J \to J+J$ scattering prevented Choi and
Santamaria from excluding majoron--neutrino couplings for the case of
light neutrino masses, $m_{\nu_\tau}\lsim 100$~eV. Third, we calculate
the majoron luminosity in the trapping regime in a different way.

Let us discuss first the mean free path of majorons.  The main source
of opacity for majorons are the processes $J+J \to\nu+\nu$, $J+\nu \to
J+\nu$, and $J+\nu \to\nu$ (cf. \cite{ch90}). Taking into account the
effective mass of the neutrinos, the corresponding mean free paths
inside the SN core with radius $r_0\sim 10$~km and density
$\rho\sim\rho_{\rm nuc}$ are given by,
\bea  \label{free1}
 l^{-1}(J+J \to\nu+\nu) &=& 1.9  \times 10^{18} g^4 
       \left(\frac{{\rm keV}}{m_{\rm eff}}\right)^2 
       \left(\frac{T}{25 \rm MeV}\right)^3  {\rm cm}^{-1}
\\
 l^{-1}(J+\nu \to J+\nu) &=& 1.5  \times 10^{18} g^4 
       \left(\frac{{\rm keV}}{m_{\rm eff}}\right)^2 
       \left(\frac{T}{25 \rm MeV}\right)^3 {\rm cm}^{-1}
\\ \label{free3}
 l^{-1}(\nu+J \to\nu) &=& 1  \times 10^{4} g^2 \; 
       \left(\frac{T}{25 \rm MeV}\right) {\rm cm}^{-1} \,.
\eea
For the effective neutrino mass inside the SN core, $m_{\rm eff}^2
\sim 2pV$, we use $m_{\rm eff} \sim 20$~keV.  Requiring that the mean
free path of the majorons is $\gsim 10$~km, we obtain that they escape
freely for $g\lsim 3 \times 10^{-6}$.

Next, we consider the luminosity for the case that majorons are not
trapped. Then they are emitted by the whole core volume with
luminosity
\bea \label{L_J}
 {\cal L}(\nu+\nu \to J+J) &=& 5.5  \times 10^{80} g^4 
                              \left(\frac{\rm keV}{m_{\rm eff}}\right)^2
 {\rm erg/s}
\\
 {\cal L}(\nu \to\nu+J) &=& 4.8 \times 10^{65} g^2 \;{\rm erg/s} \,.
\eea
We require now that the majoron luminosity during 10~s does not exceed
the maximal theoretical value of the binding energy, ${\cal L}_J \lsim
5 \times 10^{52}$erg/s, and obtain the limit $g\lsim 3 \times 10^{-7}$. In
the trapping regime, the energy loss argument therefore excludes the
band
\beq
3 \times 10^{-7} \lsim g \lsim 3 \times 10^{-6}  
\eeq
for vacuum neutrino masses $m \ll m_{\rm eff} \sim 20$~keV. 

Let us now discuss the case that majorons are trapped, $g\gsim 3
\times 10^{-6}$. We recall first the standard treatment as presented,
e.g., in Ref.~\cite{ch90}. The main assumption is that ${\cal L}_J$
can be approximated by blackbody surface emission
\beq 
 {\cal L}_J = \frac{\pi^3}{30} R_J^2 T^4
\eeq
of a {\em thermal\/} majoron--sphere which radius $R_J$ is defined to
be at optical depth 2/3,
\beq
 \int_{R_J}^\infty dr \: l^{-1}(r) = \frac{2}{3} \,.
\eeq
Assuming furthermore a temperature profile $T(r)$ outside the
supernova core, one can determine for which range of coupling
constants ${\cal L}_J$ exceeds a certain critical value.

There are two shortcomings in this argumentation. First, the profile
$T(r) \propto r^{-2}$ used for $r>r_0$ in \cite{ch90} represents the
temperature of ordinary matter (nucleons, $e^{\pm}$, photons).
However, majorons couple mainly to neutrinos and would at best be in
thermal equilibrium with neutrinos, if at all, but not with nucleons.
Second, the use of Eqs.~(\ref{free1}-\ref{free3}) which were derived
for isotropic distributions inside the core is not justified for the
highly non--isotropic distributions outside the neutrino-- or
majoron--spheres.
Using for the density $n_i(r)$ of a particle species $i$ with sphere
radius $R_i$ the expression
\beq  \label{n_escape}
 n_i(r)= \frac{{\cal L}_i }{\langle E_i\rangle} \frac{1}{4\pi r^2} \,,
\eeq
and introducing the additional factor $\langle 1-\cos\theta\rangle\sim
(R_i/r)^2$ which represents the averaging over the angle $\theta$
between the momenta of the radially outgoing test majoron and the
particle $i$ in the mean free path one finds that $ n_i(r) \propto
1/r^4$. %Thus not only the majoron--neutrino interactions outside the
%neutrino--spheres are intrinsically small, but also the neutrino
%densities are suppressed as $ n_i(r) \propto 1/r^4$. 
This renders
support to our claim that there is a sharp drop in the majoron opacity
in crossing the neutrino spheres.
Note also that it was recently stressed in Ref.~\cite{ja99} that a
naive application of the Stefan-Boltzmann law can be dangerous for the
calculation of neutrino--sphere radii.

We prefer therefore not to rely on the argumentation presented above.
Instead, we calculate the \emph{volume} rather than the \emph{surface}
luminosity, but taking into account that only majorons emitted from the
shell $[r_0-l:r_0]$ can escape \cite{be89}. We find that the majoron
luminosity drops below ${\cal L}_{\nu,{\rm tot}}\sim E_b/10$~s for $g
\gsim 2 \times 10^{-5}$.  Going even to larger values of $g$, majorons
and neutrinos are becoming so strongly coupled that it is reasonable
to assume that the SN core emits roughly the same luminosity in form
of majorons as in one neutrino species, ie. ${\cal L}_J\sim {\cal
  L}_{\nu,{\rm tot}}/6$.  Therefore, majoron emission in this regime
is not constrained at all by the energy loss argument.  Combining the
both limits obtained, majoron couplings one concludes that the range
\beq
3 \times 10^{-7} \lsim g \lsim 2 \times 10^{-5}  
\eeq
is excluded for vacuum neutrino masses $m \ll m_{\rm eff} \sim
20$~keV.

A final remark is in order. We have been using rather loosely only one
coupling constant $g$ in this section. More precisely, we mean by $g$
the element of the coupling matrix $g_{\alpha\beta}$ in the weak basis
with the largest absolute value.

\subsection{Constraints from neutrino spectra}

%%%% general:

During the Kelvin-Helmholtz cooling phase, $t\ap 1-10$~s after core
bounce, the protoneutron star slowly contracts and cools by neutrino
emission. The neutrino luminosities are governed by the energy loss of
the core and are, therefore, approximately equal for each type of
(anti-) neutrinos. Since the opacity of, e.g., $ \bar\nu_{\mu,\tau}$
is smaller than of $ \bar\nu_e$, due to their smaller cross section,
their energy-exchanging reactions already freeze out in the denser
part of the protoneutron star. Hence, one expects the spectral
temperatures of $ \bar\nu_e$ to be smaller than the one of
$\nu_h=\{\nu_{\mu,\tau}, \bar\nu_{\mu,\tau}\}$.  Typically, the
average energies $\langle E_i \rangle$ found in simulations
\cite{ja93} are
\beq
 \langle E_{\nu_e}\rangle\ap 11~{\rm MeV} \,,          \qquad
 \langle E_{ \bar\nu_e}\rangle\ap 16 \;{\rm MeV} \,,    \qquad   
 \langle E_{\nu_{h}}\rangle\ap 25~{\rm MeV} \,.
\eeq 

It is convenient to define three energy spheres $R_{E,\nu_e}$, $R_{E,
  \bar\nu_e}$ and $R_{E,\nu_{h}}$, outside of which only
energy-conserving reactions contribute to the neutrino opacity% 
\footnote{Neutrino-matter interactions are energy--dependent and,
  therefore, the concept of a neutrino--sphere is evidently an
  over-simplification.}.  These reactions, mainly neutral current
neutrino-nucleon and neutrino-nucleus scattering, do not change the
neutrino spectra, although the neutrinos still undergo several
scattering as they diffuse outward. Eventually, the neutrinos reach
the transport sphere at $R_{t,\nu_i}$, where also the
energy-conserving reactions freeze out, and escape freely. This
sphere, i.e. the surface of last scattering, is what most authors mean
with ``the'' neutrino--sphere.

For the numerical evaluation of the decay rates, we use profiles for
$\rho(r)$ and $Y_e(r)$ from Wilson's SN model \cite{SN_W} at the time
$t=6$~s after core bounce. For all radii of interest, only
anti-neutrinos can decay into neutrinos. Correspondingly, the allowed
majoron decay channels can be determined from Table~1. For the average
position of the various neutrino--spheres, we use the implicit
definition
\beq \label{spheres}
 \rho(R_{E,\nu_h}) = 2 \times 10^{13} {\rm g/cm}^3  \,,\qquad 
 \rho(R_{E, \bar\nu_e}) = 2 \times 10^{12} {\rm g/cm}^3  \,,\qquad 
 \rho(R_{E,\nu_e}) = 2 \times 10^{11} {\rm g/cm}^3  \,, 
\eeq
and
\beq
 \rho(R_{T,\nu_h}) = \rho(R_{T, \bar\nu_e}) = \rho(R_{T,\nu_e}) =
 2 \times 10^{11} {\rm g/cm}^3 \,.
\eeq

%%%%% neutrino decays:

In calculating the effect of majoron--emitting neutrino decays on the
emitted $\nu$ spectra we must distinguish three regions. If a $
\bar\nu_i$ is produced inside its own energy sphere, $r<R_{E,\bar\nu_i}$,
it is still in thermal and chemical equilibrium with the stellar
medium. Therefore, the production (or the decay) of a $ \bar\nu_i$
within $r<R_{E,\bar\nu_i}$ does not influence the emitted $ \bar\nu_i$
spectrum.  In contrast, the spectrum of $ \bar\nu_i$ neutrinos is
fixed for $r>R_{E,\bar\nu_i}$ and both its decay or its production changes
its spectra. The only difference between the two regions $R_{E,
  \bar\nu_i}<r<R_{t, \bar\nu_i}$ and $r>R_{t,\bar\nu_i}$ is the different
``effective'' velocity $v$ of the decaying neutrino. In the latter,
the neutrino can escape freely ($v=1$), while in the first it diffuses
outward with $v\ap\lambda/(R_{E, \bar\nu_i}-R_{t, \bar\nu_i})$.  Here,
$\lambda$ is the average mean free path of the neutrino $ \bar\nu_i$.
Hence, we can compute the survival probability $N( \bar\nu_i)$ of a $
\bar\nu_i$ neutrino emitted from its energy sphere as
\beq
 N( \bar\nu_i) = \exp\left\{ 
     -\int_{R_{E, \bar\nu_i}}^{R_{t, \bar\nu_i}} \frac{dr'}{v} \:
                 \Gamma_{ \bar\nu_i}(r') 
     -\int_{R_{t, \bar\nu_i}}^\infty  dr' \:\Gamma_{ \bar\nu_i}(r') 
                \right\} \,,
\eeq
where 
\beq
 \Gamma_{ \bar\nu_i} = \sum_{l=e,\mu,\tau} \Gamma( \bar\nu_i \to\nu_l +J) 
\eeq
is its total decay rate.  
Denoting with $p_{ij}[r_1,r_2]$ the probability that the decay $
\bar\nu_i \to\nu_j+J$ happens in between $r_1$ and $r_2$, it follows
that 
\beq p_{ij}[r_1,r_2] = \int_{r_1}^{r_2} \frac{dr'}{v} \: \Gamma(
\bar\nu_i \to\nu_j+J;r') N(r') \,.  \eeq
and $\sum_j p_{ij}[r_1,r_2]+N=1$, where N denotes the neutrino
survival probability.

%%%%%%%  majoron decays:

Finally, we have to consider the effect of majoron decays $J
\to\nu_i+\nu_j$ on the neutrino spectra.  If $g\lsim 3 \times
10^{-6}$, majorons escape freely. Thus they are emitted by the core
with luminosity given by Eq.~(\ref{L_J}).  For their mean energy we
assume $\langle E_J\rangle=3T\sim 50$~MeV, so that the number of
emitted majorons per unit time is $N_J={\cal L}_J/\langle E_J\rangle$. In
the intermediate regime, $3 \times 10^{-6}\lsim g \lsim 2 \times
10^{-5}$, we use the same average energy for the majorons but use the
reduced luminosity due to emission from the shell $[r_0-l:r_0]$.
Finally, for $g\gsim 2 \times 10^{-5}$, we identify simply the
majoron--sphere $R_J$ with the energy sphere of the neutrino species
which has the largest coupling to the majoron. In the following, we
will use $R_J=R_{E,\nu_h}$ and we will also assume that its average
energy is similar to the one of $\nu_h$, $\langle E_J\rangle \ap
\langle E_{\nu_{h}}\rangle\ap 25~{\rm MeV}$.

The decay probability $P=1-N$ of a majoron is given by
\beq
 N(J) = \exp\left\{ 
     -\int_{R_J}^\infty dr' \:\Gamma_{J}(r') \right\} 
\eeq
Together with the majoron luminosity ${\cal L}_J$ this allows us to
calculate the spectra of the produced neutrinos.

\subsubsection{Neutrino signal from SN1987A} 

The existing observational data on the supernova SN 1987A were already
used in Ref.~\cite{sm94}, in order to constrain the allowed
permutation between the $ \bar\nu_e$ and all other types of neutrinos
as a result of neutrino oscillations.
At 99\% (95\%) confidence level, it was found that no more than 35\%
(23\%) of the $ \bar\nu_e$ flux could be converted into
$\bar\nu_{\mu,\tau}$ or $\nu_{\mu,\tau}$. This limit can be applied to
any process which diminishes the flux of $\bar\nu_e$ and is only
weakly energy dependent, such as the case of medium-induced neutrino
decays.  Thus one can apply their results in order to restrict
the neutrino decay models under consideration here.

\begin{figure}
\centerline{\protect\hbox{
\epsfig{file=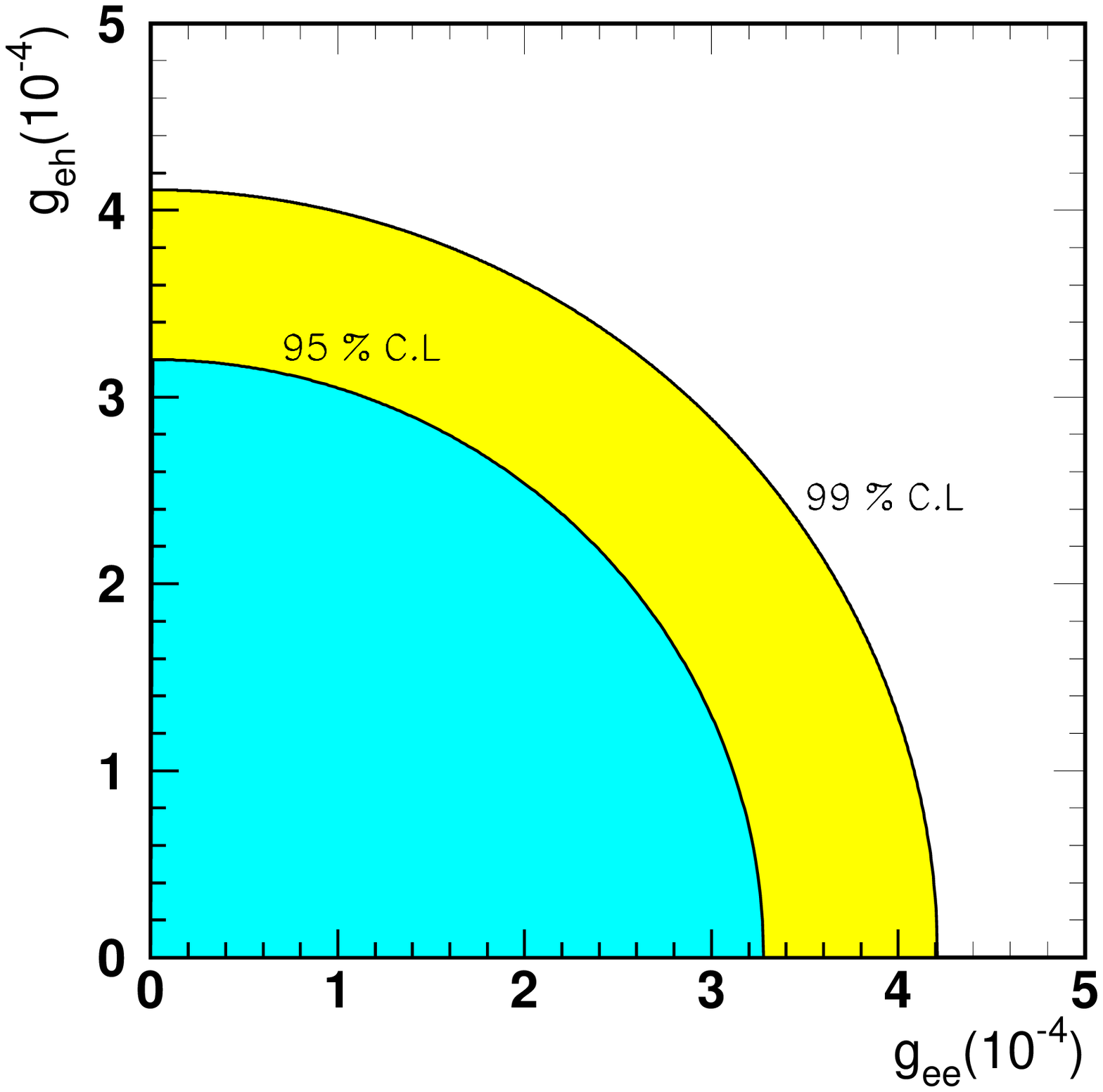,height=13.0cm,width=16.0cm}
}}
\caption{SN 1987A constraint on the majoron-neutrino effective coupling
  constants in the $g_{ee}-g_{eh}$ plane. Parameters corresponding to
  the SMA-MSW solution to the solar neutrino problem are assumed.
\label{SMA-MSW}}
\end{figure}

Since $\nu_\mu$ and $\nu_\tau$ feel the same potential, the decay
rates $\Gamma( \bar\nu_e \to\nu_\mu +J)$ and $\Gamma( \bar\nu_e
\to\nu_\tau +J)$ differ only due to their different coupling
constants. We define therefore $g_{eh}^2=g_{e\mu}^2+g_{e\tau}^2$ and
present the region excluded by the SN 1987A signal in
Fig.~\ref{SMA-MSW} in the $g_{ee}-g_{eh}$ plane.  The region excluded
at 95\% confidence level corresponds approximately to the one given by
the condition
\beq
 g_{ee}^2+g_{e\mu}^2+g_{e\tau}^2 \gsim 1 \times 10^{-7} \,.
\eeq

Up to now, we have neglected that massive neutrinos do not simply
decay but may also oscillate on their way.  In the high-density region
where the decays occur this approximation is legitimate because the
medium states coincide essentially with the weak flavour eigenstates.

However, the SN 1987A neutrinos must also propagate first through the
SN envelope, then through vacuum before they cross the Earth on the
way to the detectors.  Let us consider now what will be the impact of
neutrino oscillations for the three popular solutions the solar
neutrino problem, namely small-angle (SMA) MSW, large-angle (LMA) 
MSW and the just-so or vacuum oscillations\footnote{We follow closely
  the discussion given in  Refs.~\cite{je96,sm94}.}. 
These solutions are characterized by particular values of $\Delta_0$
and $\sin^22\theta_0$, which define different regimes in the dynamics
of neutrino propagation through the supernova.

Let us first analyze the two MSW solutions. For the usual neutrino
mass hierarchy, anti-neutrinos will not encounter an MSW resonance on
their way through the SN envelope.  Furthermore, they propagate
adiabatically, because in both cases $\Delta_0 \sim 10^{-5}$ eV$^2$.
Therefore, neutrinos that had been created as $\tilde\nu_1^+ (\rho)\ap
\bar\nu_e$ leave the star as $\tilde\nu_1^+ (\rho=0)=\nu_1^+$, i.e. as
a definite mass eigenstate.  Consequently, no neutrino oscillations
take place on the way from the SN to the Earth and the probability of
$ \bar{\nu}_e \to \bar{\nu}_\mu$ transitions (permutation factor)
equals $P_{\rm osc}=|\langle \bar{\nu}_\mu|\tilde\nu_1^+\rangle|^2 =
\sin^2\theta_0$, as long as matter effects in the Earth can be
neglected.  Since decays $ \bar\nu_\alpha \to\nu_\beta$ and
oscillations $ \bar\nu_\alpha\leftrightarrow \bar\nu_\beta$ are
decoupled, the total survival probability can be written as
\beq
 N =  N_{\rm decay} N_{\rm osc} =  N_{\rm decay} (1-P_{\rm osc}) \,.
\eeq
where $P_{\rm osc}$ is the neutrino conversion probability due to
oscillations.

In the SMA-MSW case, $\sin^22\theta_0 \approx 7 \times 10^{-3}$,
oscillations can be neglected both in the SN envelope and in the
Earth, so that $P_{\rm osc}\ap 0$. Thus, one can use directly the
results obtained above, shown in Fig.~\ref{SMA-MSW}.

\begin{figure}
\centerline{\protect\hbox{
\psfig{file=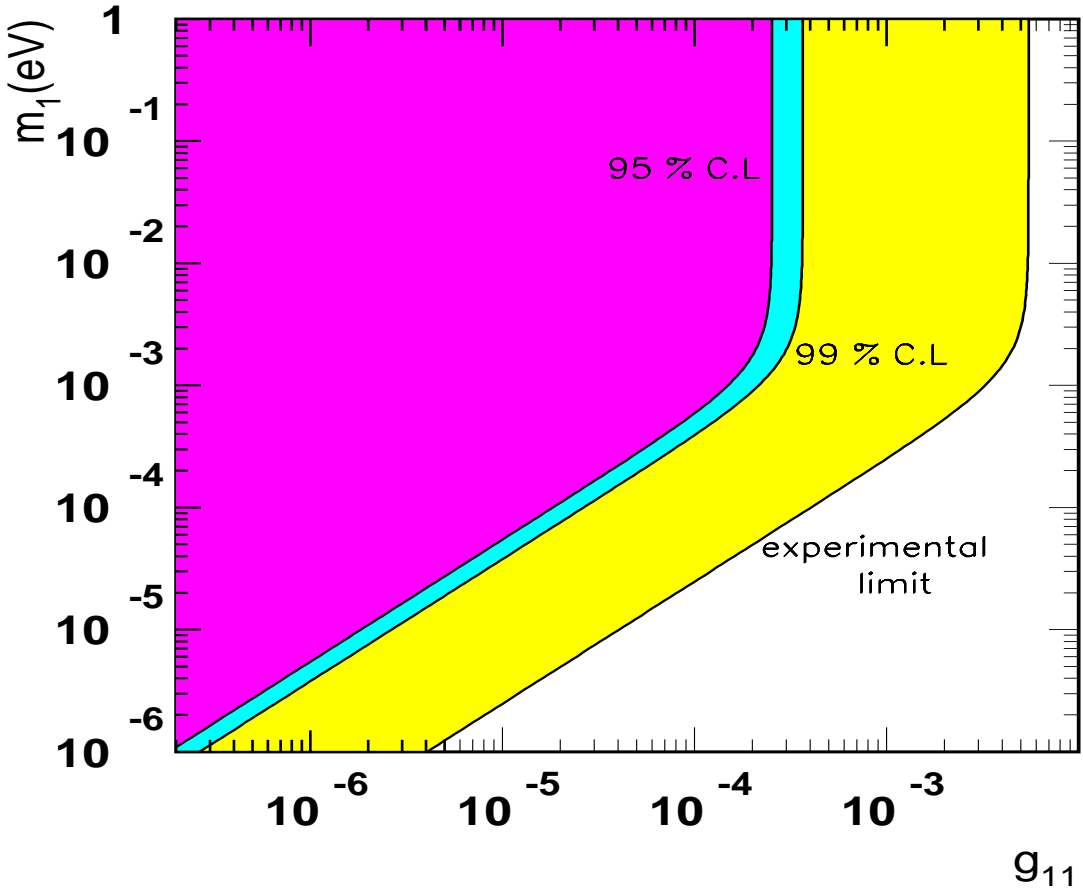,height=13.0cm,width=16.0cm}
}}
\caption{SN 1987A constraint plotted in the $m_1-g_{11}$ plane.
  Parameters corresponding to the LMA-MSW solution to the solar
  neutrino problem are assumed, $\sin^22\theta=0.6$ and
  $\Delta_0=10^{-5}$~eV$^2$.
\label{LMA-MSW}}
\end{figure}

In contrast, the matter effect inside the Earth has to be taken into
account in the LMA-MSW case, for which $\sin^22\theta_0 \approx 0.6$
\cite{valencia:1999}. The permutation probabilities due to
oscillations is given by,
\beq
 P_{\rm osc} = \sin^2 2\theta_0 
 - \sin 2\theta \sin (2\theta_0-2\theta) \sin^2 (\pi d/l_{\rm osc}) \,.
\eeq
The distance $d$ traveled inside the Earth by the neutrinos and the
average density $\rho$ are different for Kamiokande and IMB detectors.
For Kamiokande we have $d=3900$~km and $\rho=3.4$~g/cm$^3$ while for
IMB $d=8400$~km and $\rho=4.6$~g/cm$^3$. As an approximation we will
use therefore the average value $P_{\rm osc} = (12 P_{\rm osc, Kam} +
8 P_{\rm osc, IBM})/20$ according to the number of events detected in
each detector.  In Fig.~\ref{LMA-MSW}, we show the excluded region for
$\sin^2 2\theta_0=0.6$, $\Delta_0= 10^{-5}$~eV$^2$, and in addition
the experimental limit Eq.~(\ref{pi}).  The coupling $g_{22}$ is fixed
by
\beq
 g_{22}=g_{11}\sqrt{1+\frac{\Delta_0}{m_1^2}} \,.
\eeq
Therefore, for $m_1/\Delta_0 \to 0$, $g_{22}$ becomes larger for
constant $g_{11}$. Hence, the limit on $g_{11}$ becomes stronger.
Note also that we can not fix $g_{33}$ by the solar neutrino data.
Therefore, we have set conservatively $g_{33}=0$. As can be seen from
the Fig.~\ref{LMA-MSW}, for the range of masses involved, the values
$g_{11}>10^{-4}$ are excluded at $95\%$ C.L.

\begin{figure}
\centering 
 \includegraphics[height=13.0cm,width=16.0cm]{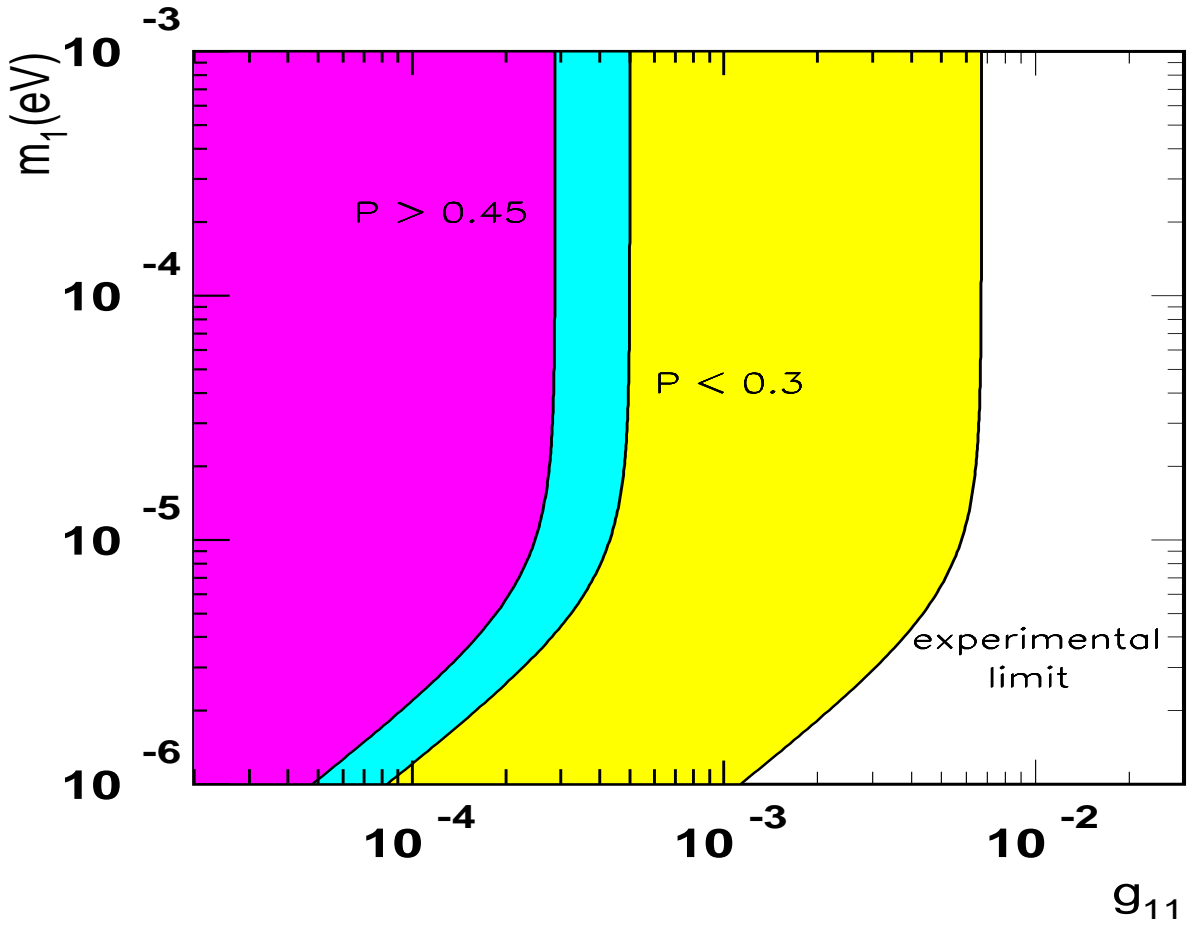}
 \caption{Regions of total transition probability 
   plotted in the $m_1-g_{11}$ plane. Here we assume the just--so
   solution to the solar neutrino problem, $\sin^22\theta=0.9$ and
   $\Delta_0=10^{-10}$~eV$^2$ and include both the effects of neutrino
   decays and oscillation.
\label{vacuum}}
\end{figure}

The remaining case to consider is the 'just-so' solution, in which
$\Delta_0 \approx 10^{-10}$~eV$^2$. For such small values of the mass
splitting, the neutrino propagation out of the SN is non-adiabatic.
Therefore, neutrinos leave the SN envelope as flavor eigenstates which
then oscillate on their way to the Earth.  Taking into account that in
this case the Earth effect is unimportant, their averaged permutation
probability due to oscillation is simply 
\beq
 P_{\rm osc} = \frac{1}{2}\; \sin^2 2\theta_0 \,.
\eeq
Let us consider now the typical value $\sin^2 2\theta_0\approx 0.9$
\cite{Barger:1999}.  Then, according to the analysis of
Ref.~\cite{sm94}, this case is already disfavored assuming only
oscillation.  In any case for completeness we have plotted in
Fig.~\ref{vacuum} the regions corresponding to $P=0.55, 0.7$ and the
experimental limit Eq.~(\ref{pi}).

To summarize, the limits we have obtained in this subsection from the
observed $ \bar\nu_e$ signal of SN 1987A are more than one order of
magnitude stronger than the limit from pion decay. They require as
theoretical input some knowledge about the spectral shape of the
emitted neutrino fluences (e.g. $\langle E_{\nu_i}\rangle$) and are
therefore to a certain extent model dependent.  However this model
dependence is much weaker than that implicit in the limit from the
collapsing phase (\ref{l1}).

\subsubsection{Constraints from a future galactic supernova}

The neutrino signal from SN 1987A observed by Kamiokande and IBM
confirmed the general astrophysical picture of a supernova explosion.
However the small number of observed neutrinos prevents a sensitive
probe of neutrino properties or of the equation of state of
super--dense matter. Meanwhile, several new experiments like
Super--Kamiokande or the Sudbury Neutrino Observatory (SNO) have been
constructed and it seems therefore interesting to estimate their
sensitivity to new physics, such as the majoron--emitting neutrino
decays, through their ability to determine the spectra of SN
neutrinos.

All detectors in the near future can detect SN only in our own Galaxy
and in the Large and Small Magellanic Cloud, with however a much
smaller rate than expected for the Milky Way. Therefore, we use in the
following 10 kpc as the reference distance to the hypothetical
Supernova.  Then, according to Ref.~\cite{bu92}, the number of $
\bar\nu_e+p \to n+e^+$ events expected in SK is 5310, compared to 220
events in all other channels. While a light-water detector like SK
acts like a $ \bar\nu_e$ filter, a heavy-water detector has a broad
response to all types of neutrinos: from the total of 781 events
expected at SNO, 200 events are neutral-current reactions of $\nu_h$
(=$\nu_\mu, \bar\nu_\mu,\nu_\tau$ or $ \bar\nu_\tau$) with deuterium.
In addition, there are 82 charged-current reactions of $\nu_e$ with
deuterium. The combined spectral information of these reactions should
contain the major part of the information extractable from all data.
In the following, we will consider therefore only the reaction $
\bar\nu_e+p \to n+e^+$ for SK, and $\nu_h+d \to\nu_h+n+p$ and $\nu_e+d
\to p+p+e^-$ for SNO.

The predicted number of events has a rather strong dependence on how
the cooling of the SN is modeled. In Ref.~\cite{be99}, e.g., the
expected number of $\nu_h+d \to\nu_h+n+p$ reactions is 485 in SNO.  An
important question is therefore how well different astrophysical
and/or particle physics processes (e.g. neutrino oscillations versus
decays) can be disentangled using observational data. We will comment
shortly on this question at the end of this section.

The neutrino spectrum $n_\nu(E_\nu)$ emitted by the SN can be be
inferred in a rather indirect way from the experimentally measured
energies $E_i$ of the observed events.  Let us consider first the
reaction $ \bar\nu_e+p \to n+e^+$ and $\nu_e+d \to p+p+e^-$.
Experimentally, the neutrino energy $E_\nu$ is reconstructed in both
reactions from the number of photo-multipliers that have been
triggered by Cherenkov photons emitted by the $e^\pm$.  The
probability that a certain energy $E_i$ is ascribed to an event $i$
where the $e^\pm$ has the energy $E_e$ depends in principle on the
geometry and the details of the detector.  For our purposes it is
enough to take into account the Poissonian nature of the emission and
detection of the Cherenkov photons. In this case the probability
distribution function that the energy $E_i$ is ascribed to an $e^\pm$
with energy $E_e$ is a Gaussian distribution,
\beq \label{G}
 G(E_i,E_e) = \frac{1}{\sqrt{2\pi}\sigma(E_e)} \exp\left(-
          \frac{ (E_e-E_i)^2 }{2\sigma^2 (E_e)} \right) 
\eeq                     
with $\sigma(E_e)=\sqrt{E_e E_\sigma}$, where $E_\sigma$ is the energy
resolution.  Following Refs.~\cite{det}, we use $E_\sigma=0.22$~MeV
for SK and $E_\sigma=0.20$~MeV for SNO.  Apart from its resolution
$E_\sigma$, the other main feature of a detector is its efficiency
$\epsilon(E)$. For both experiments, we use the approximation
$\epsilon(E)=\epsilon_0\theta(E-E_{\rm cut})$ with $\epsilon_0=1$ and
a conservative value for $E_{\rm cut}=7$~MeV for SK and $E_{\rm
  cut}=5$~MeV for SNO.  Combining the Gaussian kernel $G(E_i,E_e)$
with the detector efficiency $\epsilon(E)$, the relation between the
$e^\pm$ energy $E_e$ and the measured energy $E_i$ is
\beq  \label{data}
 n(E_i) = \epsilon_0\int_{E_{cut}}^\infty dE \; G(E_i,E_e) n_e(E_e) \:.
\eeq
In contrast to the two reactions considered above, the neutral-current
reaction $\nu_h+d \to\nu_h+n+p$ used to detect $\nu_h$ in SNO does not
allow to measure the energy $E_\nu$.  Nevertheless, it is possible to
reconstruct partially the $\nu_h$ energy spectrum comparing the
observed signal in different detector materials with different energy
responses (cf., e.g., Fig.~3 of Ref.~\cite{be99}).
 
As next ingredient, we need the relation between the time--integrated
neutrino spectra $n_i(E)$ and the spectra of the secondary $e^\pm$'s.
For the reaction $ \bar\nu_e+p \to n+e^+$, the positron spectrum is
given by
\beq \label{pos}
 n_e(E_e) = \frac{N_p}{4\pi D^2} \sigma_{ \bar\nu_e+p \to n+e^+} (E+\mu) 
                        n_{ \bar\nu_e} (E+\mu) \,
\eeq
where $N_p$ is the number of target protons in SK and $D$ the distance
to the supernova. For the cross section, we use
\beq
 \sigma_{ \bar\nu_e+p \to n+e^+} = \sigma_0 E^2 \left( 1-\frac{\mu}{E} \right) 
    \left( 1-\frac{2\mu}{E}+\frac{\mu^2+m_e^2}{E^2} \right)^{1/2}
\eeq
with $\sigma_0=2.3 \times 10^{-44}$~cm$^2$, where $ \mu=1.227$~MeV is
the proton-neutron mass difference and $m_e$ the electron mass. In the
case of the two neutrino-deuterium reactions, we use the
cross-sections tabulated in Ref.~\cite{sigma} and replace $N_p$ by the
number of deuterium nuclei.

Numerical simulations of the neutrino transport show that the
instantaneous neutrino spectra can be described as Fermi-Dirac
distributions with an effective degeneracy parameters $\eta_i$
\cite{ja89,gi89}.  The instantaneous neutrino spectra found are
pinched, i.e. their low- and high-energy parts are suppressed relative
to a Maxwell-Boltzmann distribution. Janka and Hillebrandt found that
during the cooling process the effect of pinching becomes less
important. Moreover, the pinching of the instantaneous neutrino
spectra is compensated by the superposition of different spectra with
decreasing temperatures.

In the following, we assume for the time-integrated energy spectra of
the different neutrino types
\beq  \label{n}
 n_{i} (E) = N_i \frac{E^2}{e^{E/T_i-\eta_i}+1}
\eeq
with $\langle E_{\nu_e}\rangle=11$~MeV,
$\langle E_{ \bar\nu_e}\rangle= 16$~MeV and
$\langle E_{\nu_h}\rangle=25$~MeV.
For the degeneracy parameter, we use the values of the lower end of the
range found by Janka and Hillebrandt, namely $\eta=3$ for $\nu_e$,
$\eta=2$ for $ \bar\nu_e$ and $\eta=0$ for $\nu_h$. Then the relation
between the effective temperature $T_i$ is given by
$3T\ap0.751\langle E_{\nu_e}\rangle$ for $\nu_e$,
$3T\ap 0.832\langle E_{ \bar\nu_e}\rangle$ for $ \bar\nu_e$ and
$3T=\langle E_{\nu_h}\rangle$ for $\nu_h$.

\begin{figure}
\centering 
 \includegraphics[height=13.0cm,width=16.0cm]{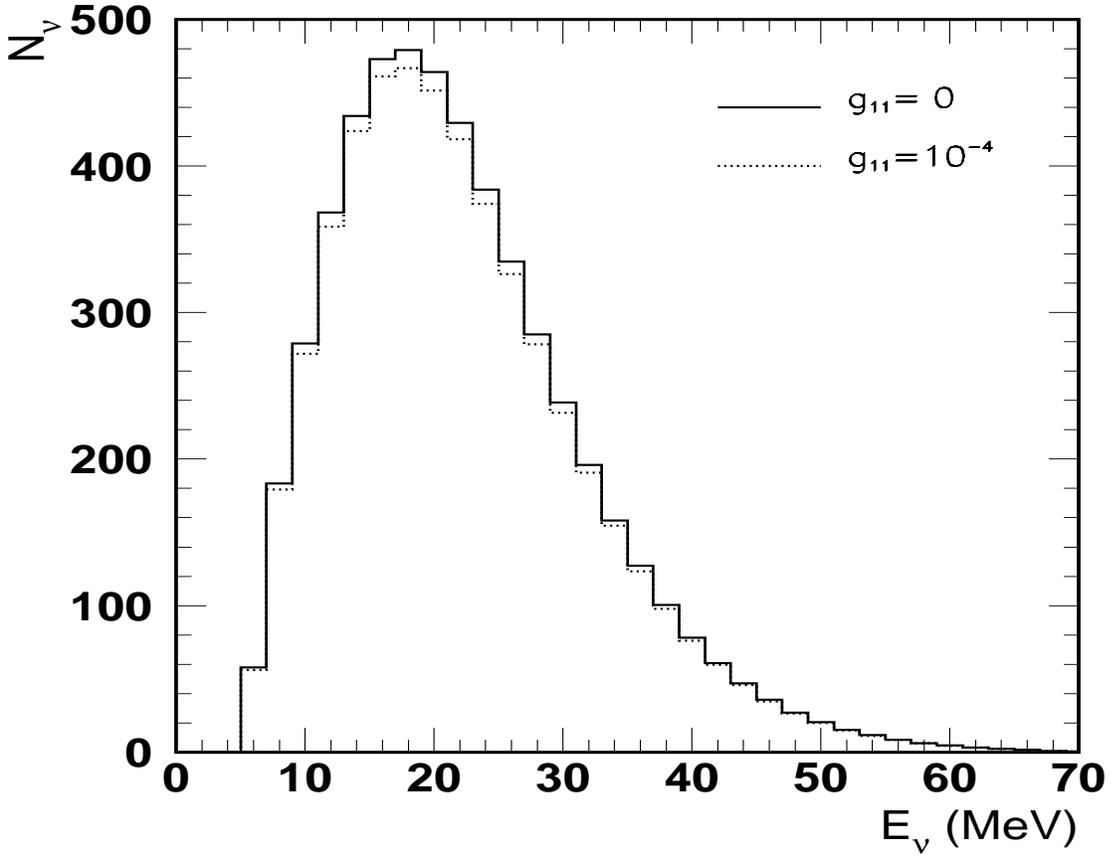}
 \caption{Signal expected in Super-Kamiokande due to the reaction 
   $ \bar \nu_e + p \to n+e^+$ for a galactic supernova.  Here $N_\nu$
   denotes the number of events for $g_{11}=0$ (solid line) and
   $g_{11}=10^{-4}$ (dashed line) with $g_{11}=g_{22}=g_{33}$.
\label{eq1.eps}}
\end{figure}

\begin{figure}
\centering 
 \includegraphics[height=13.0cm,width=16.0cm]{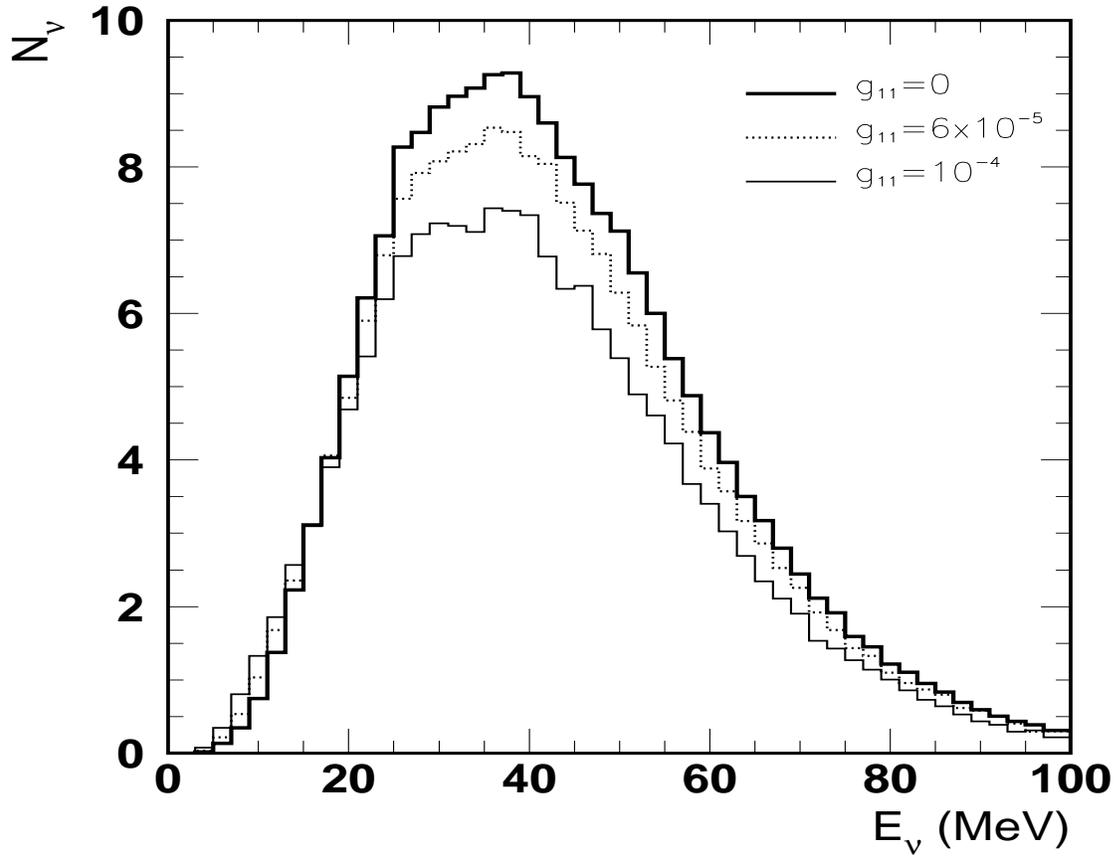}
 \caption{Signal expected at SNO due to the reaction $\nu_h + D \to \nu_h+p+n$
   for a galactic supernova. Here $N_\nu$ denotes the number of events
   for the indicated values of $g_{11}$ with $g_{11}=g_{22}=g_{33}$.
\label{eq3.eps}}
\end{figure}

We have performed a simulation of the expected neutrino signals in SK
and SNO in the case of a galactic Supernova. Here have assumed
$g_{11}\ap g_{22}\ap g_{33}$, as suggested by a scheme with all three
neutrino masses are quasi-degenerate, in the eV range~\cite{DEG}.
Moreover, here we have chosen the solar and atmospheric mixing angles
to be maximal~\cite{bimax98}.  The resulting average neutrino signals
disregarding neutrino oscillations are shown for several values of
$g_{11}$ in Fig.~\ref{eq1.eps} and \ref{eq3.eps}.  As one could expect
from the discussion in Section~3.3.1, for the chosen values of $g$
neutrino decays do not influence drastically the $ \bar\nu_e$ signal
(Fig.~\ref{eq1.eps}).  In contrast the $\nu_h$ signal shown in
Fig.~\ref{eq3.eps} shows two main features.  First, in the presence of
majoron interactions, the spectra show a surplus of low-energy
($E\lsim 18$~MeV) and a deficit of high-energy ($E\gsim 18$~MeV)
$\nu_h$ compared to the reference Standard Model spectrum with $g=0$.
One sees that the decays $ \bar\nu_h \to \nu_l+J$ reduce the $\nu_h$
fluence by at most a factor 2.  Since the decays $\bar\nu_h \to
\nu_l+J$ and the majoron decays $J \to\nu+\nu'$ both produce also
low-energy $\nu_{\tau,\mu}$, the $\nu_h$ signal shows a surplus of
events at low energies.

We have not shown the $\nu_e$ spectra because they are basically
unchanged with respect to the reference spectrum with $g=0$. Since the
$\nu_e$ energy sphere at $R_{E,\nu_e}$ is located at a much smaller
density than those of $ \bar\nu_e$ and $\nu_h$, most ($\sim 95\%$) $
\bar\nu_e$ and $\nu_h$ decay inside $R_{E,\nu_e}$ where the $\nu_e$
are still in chemical equilibrium. Therefore, its spectrum is hardly
affected by $ \bar\nu_e$ and $\nu_h$ decays.

Finally, we want to comment on the dependence of our results on the
assumed astrophysical parameters. First note that, an overall
suppression of all three neutrino signals could be explained more
naturally by an astrophysical reason than by neutrino decays. Similarly
if a suppression with respect to the expectations of a given SN model
were observed only in the $\nu_h$ signal, this would indicate an
astrophysical explanation, since the $\nu_h$ temperature has the
largest uncertainty.  It is therefore of importance that part of the
spectral information of the $\nu_h$ signal can be recovered by
comparing the signal in different detector materials.  The signature
for majoron neutrino decays would be two-fold. First, a reduction of
the observed $\nu_h$ fluence compared to the one expected from the
observed $\nu_h$ temperature. Second, a non-thermal $\nu_h$ spectrum
with a surplus of low-energy and a deficit of high-energy $\nu_h$.

A more quantitative study of the restrictions on the majoron--neutrino
couplings attainable at SK and SNO would require a detailed likelihood
analysis. However, from Fig.~\ref{eq3.eps} one expects that the
sensitivity of these experiments will be restricted to $g \lsim 5
\times 10^{-5}$. Therefore, we conclude that although these
experiments could narrow down the allowed window for majoron--neutrino
interactions, but not close it completely.

\section{Conclusions}

We have reconsidered the influence of majoron neutrino decays on the
neutrino signal of supernovae in the full range of allowed neutrino
masses, in the light of recent Super--Kamiokande data on solar and
atmospheric neutrinos.  In the high--density supernova medium the
effects of Majoron--emitting neutrino decays are important even if
they are suppressed \emph{in vacuo} by small neutrino masses and/or
off-diagonal couplings. In contrast to previous works, we have
considered scattering and decay processes, taking into account medium
effects for both kinds of processes. The dispersive effects of the
dense SN core are particularly important since the currently favoured
interpretation of the solar and atmospheric neutrino data points
towards light neutrinos.

The observation of SN 1987A excludes two parts of the possible range
of neutrino majoron coupling constants. In the range $3 \times 10^{-7}
\lsim g \lsim 2 \times 10^{-5}$, where $g$ is the largest element of
the coupling matrix $g_{\alpha\beta}$, the supernova looses to much
energy into majorons, thereby shortening the neutrino signal too much.
For larger couplings, the fluence of escaping $ \bar\nu_e$ is reduced
due to decays $ \bar\nu_e \to\nu_l+J$. Depending on which is the
correct solution to the solar neutrino problem, different values of
$g_{\rm eff}$ can be excluded: $g_{ee}^2+g_{eh}^2\gsim 1 \times
10^{-7}$ for SMA-MSW (Fig.~\ref{SMA-MSW}) and $g_{11}\gsim 1 \times
10^{-4}$ for LMA-MSW (Fig.~\ref{LMA-MSW}).  In the case of vacuum
oscillations, the predicted number of $ \bar\nu_e$ events is already
disfavored (even in the absence of decays) by the observed number of
events and by the current theoretical understanding of supernova
explosions. The corresponding sensitivities are displayed in
Fig.~\ref{vacuum}.

Finally, we have discussed the potential of Superkamiokande and the
Sudbury Neutrino Observatory to detect majoron neutrino interactions
in the case of a future galactic supernova. We have found that
although these experiments could narrow down the allowed window for
majoron--neutrino interactions, but not close it completely, reaching
a sensitivity at the few $\times 10^{-5}$ level, as seen from
Fig.~\ref{eq3.eps}.

%%%%%%%%%%%%%%%%%%%%%%%%%%%%%%%%%%%%%%%%%%%%%%%%%%%%%%%%%%%%%%%%%%%%%%%%%%%%%%%
\section*{Acknowledgements}

We are grateful to J.F.~Beacom, Z.~Berezhiani, J.A.~Pons and A.~Rossi
for helpful comments. We would like to thank especially H.-T. Janka for
correspondence and for sending his supernova simulation data. This
work was supported by DGICYT grant PB95-1077 and by the EEC under the
TMR contract ERBFMRX-CT96-0090.  MK has been supported by a
Marie-Curie grant and RT by a fellowship from Generalitat Valenciana.

%%%%%%%%%%%%%%%%%%%%%%%%%%%%%%%%%%%%%%%%%%%%%%%%%%%%%%%%%%%%%%%%%%%%%%%%%%%%%%%

%%%%%%%%%%%%%%%%%%%%%%%%%%%%%%%%%%%%%%%%%%%%%%%%%%%%%%%%%%

\newpage


\end{document}